\documentclass[aip,prb,twocolumn,showpacs]{revtex4}

\usepackage{graphics,graphicx,amssymb,amsmath,color}
\begin{document}

\title{Magnetic order of the hexagonal rare-earth manganite Dy$_{0.5}$Y$_{0.5}$MnO$_{3}$}
\author{Joel~S.~Helton$^{1,*}$}
\author{Deepak K. Singh$^{1,2}$}
\author{Harikrishnan S. Nair$^{3}$}
\author{Suja Elizabeth$^{4}$}

\affiliation{$^{1}$NIST Center for Neutron Research, National Institute of Standards and Technology,
Gaithersburg, MD 20899, USA}
\affiliation{$^{2}$Department of Materials Science and Engineering, University of Maryland, College Park, MD 20742, USA}
\affiliation{$^{3}$J\"{u}lich Centre for Neutron Science, Forschungszentrum J\"{u}lich, Outstation at FRM-II, D-85747 Garching, Germany}
\affiliation{$^{4}$Department of Physics, Indian Institute of Science, Bangalore 560012, India}
\date{\today}
\begin{abstract}
Hexagonal Dy$_{0.5}$Y$_{0.5}$MnO$_{3}$, a multiferroic rare-earth manganite with geometrically frustrated antiferromagnetism, has been investigated with single-crystal neutron diffraction measurements.  Below 3.4~K magnetic order is observed on both the Mn (antiferromagnetic) and Dy (ferrimagnetic) sublattices that is identical to that of undiluted hexagonal DyMnO$_{3}$ at low temperature.  The Mn moments undergo a spin reorientation transition between 3.4~K and 10~K, with antiferromagnetic order of the Mn sublattice persisting up to 70~K; the antiferromagnetic order in this phase is distinct from that observed in undiluted (h)DyMnO$_{3}$, yielding a qualitatively new phase diagram not seen in other hexagonal rare-earth manganites.  A magnetic field applied parallel to the crystallographic $c$ axis will drive a transition from the antiferromagnetic phase into the low-temperature ferrimagnetic phase with little hysteresis.
\end{abstract}
\maketitle

\section{INTRODUCTION}

The crystalline structure of rare-earth manganites ($R$MnO$_{3}$ with $R$~=~Y, Sc, or a lanthanide) is determined by the ionic radius of the $R^{3+}$ cation.  Materials with a large $R^{3+}$ ionic radius ($R$~=~La through Tb) crystallize in an orthorhombic perovskite structure, while materials with a smaller $R^{3+}$ ionic radius ($R$~=~Y, Sc,  or Ho through Lu)\cite{Froehlich,Munoz2000,Tomuta,Sugie,Munoz,Yen} crystallize in a hexagonal structure.  DyMnO$_{3}$ typically crystallizes in the orthorhombic structure,\cite{Prokhnenko} but with proper growth conditions hexagonal (h)DyMnO$_{3}$ can be stabilized.\cite{Ivanov2006,Harikrishnan2009}  The hexagonal rare-earth manganites, (h)$R$MnO$_{3}$, are paraelectric at very high temperatures but display a structural transition ($T_{C}$~$\approx$~1000~K) to a ferroelectric phase with the noncentrosymmetric $P6_{3}cm$ space group.\cite{VanAken}  The (h)$R$MnO$_{3}$ materials feature slightly distorted triangular lattice planes of Mn ions; the antiferromagnetic nearest-neighbor exchange interaction leads to geometrically frustrated magnetism.  Below a N\'{e}el temperature of $\approx$100~K these materials are magnetically ordered, with easy-plane antiferromagnetic order of the Mn sublattice coexisting with the ferroelectric order.  Many hexagonal rare-earth manganites display one or more spin reorientation transitions of the Mn moments at lower temperatures\cite{Vajk2005,Lonkai} or under an applied field;\cite{Lorenz,Vajk2006} the $R$ ions also form distorted triangular lattice planes and, for magnetic ions, will order along the $c$ axis.  Several of these materials have attracted interest because of significant magnetoelectric\cite{Lottermoser,Ueland} or magnetoelastic\cite{Lee} effects.  Despite the structural similarities between the various members of the (h)$R$MnO$_{3}$ family, the magnetically ordered structures often differ, with structures that transform according to each of the four one-dimensional irreducible representations of the point group observed in at least one material.  The dependence of the magnetic ordering of the Mn sublattice on the rare-earth element has been attributed to the ionic radius of the $R^{3+}$ cation\cite{Kozlenko} or a biquadratic 3$d$-4$f$ magnetic coupling.\cite{Wehrenfennig}

Hexagonal (h)DyMnO$_{3}$, with the largest rare-earth ionic radius of this series, features an interesting magnetic phase diagram.\cite{Harikrishnan2009,NandiDMO,Wehrenfennig}  In the low-temperature phase (below $\approx$8~K) both the Mn and Dy sublattices are magnetically ordered according to the $\Gamma_{2}$ irreducible representation.  At a temperature of~8~K the Mn moments undergo a spin reorientation transition and are ordered in the $\Gamma_{4}$ representation up to 68~K.\cite{Wehrenfennig}  X-ray resonant magnetic scattering measurements also find a weak Dy moment in this temperature range, ordered according to the $\Gamma_{3}$ representation.\cite{NandiDMO}  This incompatible order, with different irreducible representations present on the two sublattices, calls into question long standing assumptions about the rigidity of the 3$d$-4$f$ interaction in (h)$R$MnO$_{3}$ materials.\cite{Wehrenfennig}  YMnO$_{3}$ features magnetism only on the Mn sublattice and is known to order in the $\Gamma_{3}$ irreducible representation below $T_{N}$~$\approx$~75~K.\cite{Fiebig2000}  YMnO$_{3}$ has also been reported to feature large magnetoelastic effects\cite{Lee} and coupling between electric and magnetic domains.\cite{Fiebig2002}  Compounds where magnetic rare-earth ions are partially replaced with nonmagnetic Y$^{3+}$ ions, such as Ho$_{1-x}$Y$_{x}$MnO$_{3}$\cite{Zhou,Vasic} and Er$_{1-x}$Y$_{x}$MnO$_{3}$,\cite{Sekhar,Vajk2011}  allow for further examination of the role that the rare-earth ions play in determining the magnetic structure and open up the possibility of novel phase diagrams not observed in undoped compounds.  We report single-crystal neutron diffraction studies of hexagonal Dy$_{0.5}$Y$_{0.5}$MnO$_{3}$ (DYMO) and find evidence for a spin reorientation transition not seen in other hexagonal rare-earth manganites at zero field.

\section{EXPERIMENT}

Large, high-quality single-crystal samples of Dy$_{0.5}$Y$_{0.5}$MnO$_{3}$ were prepared as previously reported.\cite{Harikrishnan2011}  DYMO crystallizes in the hexagonal $P6_{3}cm$ space group (\#185) with lattice parameters (at 300~K) of $a$~=~$b$~=~6.161(1)~{\AA} and $c$~=~11.446(2)~{\AA}.  As in other hexagonal rare-earth manganites, Mn$^{3+}$ ions ($S$~=~2) occupy the $6c$ positions at ($x$,~0,~0) with $x$~$\approx$~$\frac{1}{3}$.  A material with $x$~=~$\frac{1}{3}$ would feature perfect triangular lattice planes; in DYMO $x$~=~0.3379(4) yielding a slightly distorted triangular lattice.  The rare-earth $R^{3+}$ ions occupy two crystallographically distinct sites, at the $2a$ and 4$b$ Wyckoff positions, and also form distorted triangular lattice planes.  In DYMO the rare-earth sites are occupied with equal probability by nonmagnetic Y$^{3+}$ ions and $^{6}H_{15/2}$ Dy$^{3+}$ ions ($gJ$~=~10~$\mu_{B}$), as determined by x-ray powder diffraction on crushed single crystals.\cite{Harikrishnan2011}  Previously reported specific heat measurements found peaks at 3~K and 68~K; analogously with other hexagonal rare-earth manganites such as DyMnO$_{3}$ it was suggested that these peaks correspond to the onset of antiferromagnetic order of the Mn lattice, $T_{N}^{\textrm{Mn}}$~$\approx$~68~K, and ferrimagnetic order of Dy moments on the rare-earth lattice, $T_{N}^{\textrm{Dy}}$~$\approx$~3~K.

Neutron diffraction experiments were carried out using the BT9 thermal triple-axis spectrometer at the NIST Center for Neutron Research.  The neutron initial and final energies were selected using the (0~0~2) reflection of the pyrolytic graphite (PG) monochromator and analyzer.  We used 40$^{\prime}$-47$^{\prime}$-40$^{\prime}$-open collimation as well as PG filters to reduce contamination of the beam with higher-order neutron wavelengths.  The detailed temperature dependence of four magnetic reflections (shown in Fig.~\ref{PeaksTemp}) was measured using a large ($\approx$1~g) single crystal mounted in the ($H$~0~$L$) scattering plane at a fixed neutron energy of 30.5~meV ($\lambda$~=~1.64~{\AA}).  Refinement of the ordered moments utilized diffraction measurements taken at temperatures of 1.6~K, 25~K, and 120~K at a fixed neutron energy of 30.5~meV.  In order to minimize absorption of neutrons by the sample, these measurements were taken with small single crystals: an 8-mg sample mounted in the ($H$~0~$L$) scattering plane and a 6-mg sample mounted in the ($H$~$K$~0) scattering plane.  Some magnetic reflections also featured weak nuclear contributions which were removed by subtracting the 120~K intensity; the Debye-Waller factor has been ignored, which is justifiable, given that the intensities of nuclear Bragg peaks with no magnetic intensity [such as (0~0~$L$) where $L$ is even] remained constant within the statistical uncertainties between 1.6~K and 120~K.  The scattering intensity in absolute units, and from this the value of the ordered moments, was determined by normalizing the intensities of nuclear Bragg peaks measured at 120~K to the calculated nuclear intensities.  The intensities of the (1~0~0) and (3~$\bar{1}$~0) reflections in a magnetic field were measured on the 6-mg sample mounted in the ($H$~$K$~0) scattering plane; the sample was placed inside a helium flow dewar with a minimum temperature of 4~K inserted into a 7-T vertical field superconducting magnet.  These data were taken at a fixed neutron energy of 14.7~meV ($\lambda$~=~2.36~{\AA}).  All neutron diffraction data are reported in terms of the integrated intensity, integrated over a rocking curve ($\theta$ scan) through the peak position.

The allowed magnetic structures in DYMO correspond to the six irreducible representations of the $P6_{3}cm$ space group with propagation vector $\vec{k}$~=~0; as in other hexagonal rare-earth manganites, only the four one-dimensional irreducible representations (designated as $\Gamma_{1}$ through $\Gamma_{4}$) are required to describe the observed structures.\cite{Sikora,Munawar}  For each of these representations the Mn sublattice displays antiferromagnetic order within the $ab$ plane, with the three spins around any triangle oriented 120$^{\circ}$ apart.  The $R^{3+}$ sublattice is ordered along the crystallographic $c$ axis.  The rare-earth 4$b$ sites are antiferromagnetically ordered for the $\Gamma_{1}$, $\Gamma_{3}$, and $\Gamma_{4}$ representations; the 2$a$ rare-earth sites are antiferromagnetically ordered in the $\Gamma_{3}$ representation and paramagnetic for the $\Gamma_{1}$ and $\Gamma_{4}$ representations.  The 2$a$ and 4$b$ sites are each ferromagnetically ordered in the $\Gamma_{2}$ representation, with the coupling between the two sites yielding either ferromagnetism or ferrimagnetism.  In phases where only the Mn sublattice is magnetically ordered a hexagonal rare-earth manganite with $x$~=~$\frac{1}{3}$ (featuring undistorted triangular lattice planes) would have perfectly identical neutron scattering structure factors in either the $\Gamma_{1}$ or $\Gamma_{3}$ representations, as well as in the $\Gamma_{2}$ or $\Gamma_{4}$ representations.  Unambiguous differentiation between these homometric magnetic structures requires complementary measurement techniques such as optical second harmonic spectroscopy.\cite{Froehlich,Fiebig,Fiebig2000}  The distortion of the triangular planes leads to some variation in the neutron scattering structure factors; however, these results, from single-crystal diffraction measurements on a strongly absorbing sample, will not be able to distinguish between the $\Gamma_{1}$ and $\Gamma_{3}$ structures.

\section{RESULTS}

\subsection{Zero-field magnetic order}

\begin{figure}
\centering
\includegraphics[width=9.3cm] {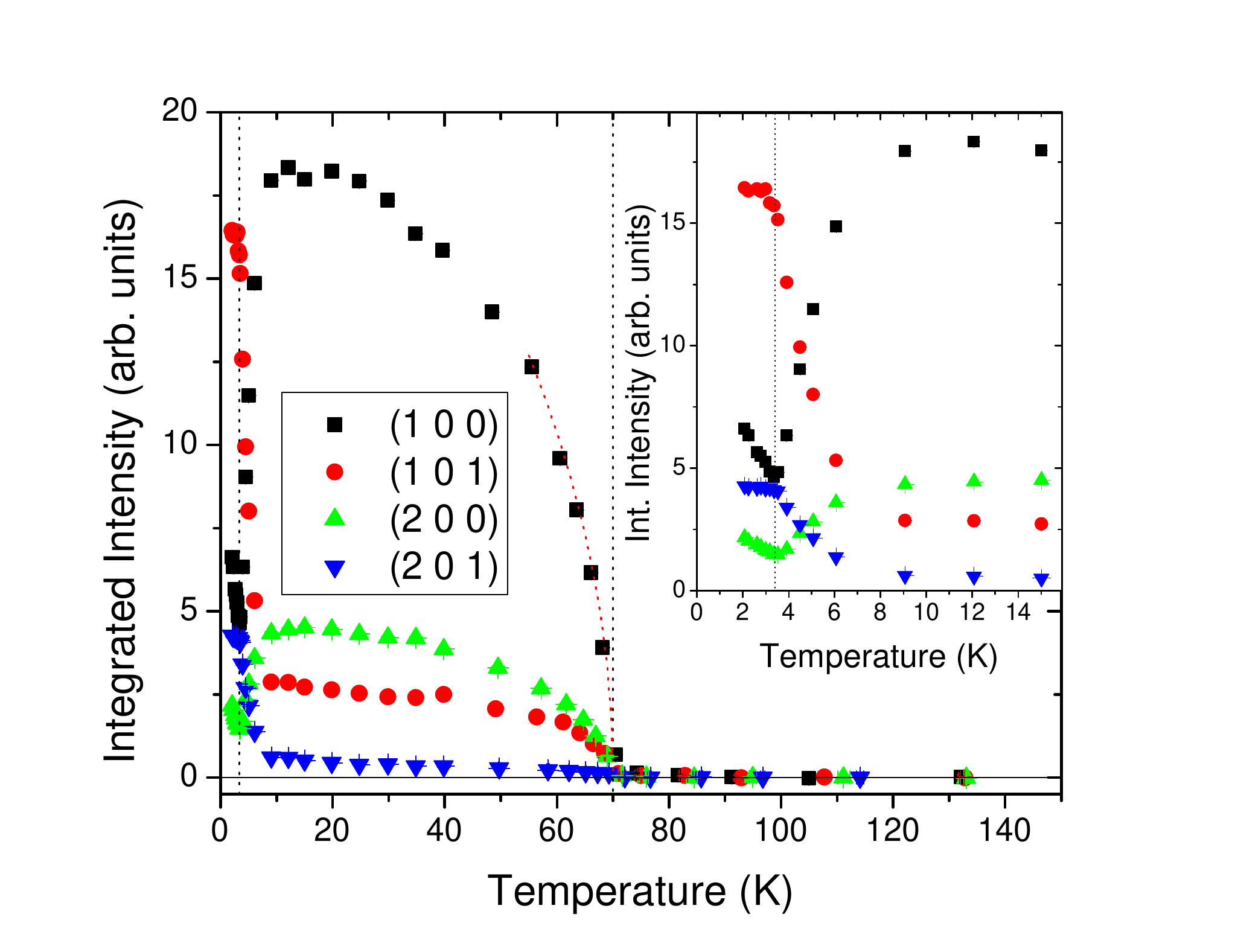} \vspace{-6mm}
\caption{(Color online) Integrated magnetic intensities of the (1~0~0), (1~0~1), (2~0~0), and (2~0~1) Bragg reflections as a function of temperature.  All data were measured while warming.  The inset shows a closer view of the low temperature portion of the graph.  Transitions are observed at $T_{N}^{\textrm{Mn}}$~=~70~K and $T_{N}^{\textrm{Dy}}$~=~3.4~K; these temperatures are designated with dashed vertical lines.  The red dotted line reflects a fit of the (1~0~0) intensity to an order parameter form with $\beta$~=~0.25~$\pm$~0.05.  Error bars throughout this article are statistical in nature and represent one standard deviation.}
\label{PeaksTemp}
\end{figure}

Figure~\ref{PeaksTemp} displays the temperature dependence of the integrated intensities of the (1~0~0), (1~0~1), (2~0~0), and (2~0~1) Bragg reflections measured while warming from 2.1~K to 130~K.  The background and any high temperature nuclear contributions to the intensities have been subtracted.  At 2.1~K all four reflections display magnetic intensity, with the (1~0~1) reflection the strongest.  Between 2.1~K and 3.4~K the measured intensities of the ($H$~0~1) reflections remain relatively constant while those of the ($H$~0~0) reflections steadily decrease with increasing temperature, with an intensity at 3.4~K that is about 68\% of the base temperature intensity.  On increasing the temperature beyond 3.4~K, the intensities of the ($H$~0~0) peaks steadily increase while the intensities of the ($H$~0~1) peaks decrease so that the (1~0~0) reflection becomes the strongest measured; this rapid change in intensity persists only to around 10~K.  This will be shown to correspond to a spin reorientation transition of the Mn moments and perhaps explains the 10~K anomaly reported\cite{Harikrishnan2011} in the derivative of 1/$\chi$.  Above 10~K the intensities of all peaks slowly decrease until reaching zero at 70~K. The intensity can be fit to an order parameter form: $I(T) \, \propto \, (|T-T_{N}|/T_{N})^{2\beta}$.  The dotted red line in Fig.~\ref{PeaksTemp} is a fit of the (1~0~0) intensity to this form with $\beta$~=~0.25~$\pm$~0.05.

\begin{table}
\caption{Calculated magnetic scattering intensities for the reflections displayed in Fig.~\ref{PeaksTemp}.  The most intense reflection in each column has been normalized to 100.  In the low-temperature phase below 3.4~K the data are consistent with $\Gamma_{2}$ order for both the Mn and Dy sublattice.  In the antiferromagnetic phase between 10~K and 70~K the data are consistent with Mn sublattice order in either the $\Gamma_{1}$ or $\Gamma_{3}$ irreducible representation.}
\vspace{5pt}
\begin{tabular}{ c  c  c  c  c  c  c  c }
\hline \hline
\multicolumn{2}{c}{ } & \multicolumn{4}{c}{Mn order} & \multicolumn{1}{c}{ } & \multicolumn{1}{c}{Dy order}\\
\hline
& \hspace{5pt} & \hspace{4pt} $\Gamma_{1}$ \hspace{4pt} & \hspace{4pt} $\Gamma_{2}$ \hspace{4pt} & \hspace{4pt} $\Gamma_{3}$ \hspace{4pt} & \hspace{4pt} $\Gamma_{4}$ \hspace{4pt} & \hspace{10pt} & \hspace{4pt} $\Gamma_{2}$ \hspace{4pt}\\
\hline
\hspace{2pt} (1 0 0) \hspace{2pt} & & 100 & 0 & 100 & 0 & & 100 \\
(1 0 1) & & 17 & 100 & 15 & 100 & & 0 \\
(2 0 0) & & 32 & 0 & 27 & 0 & & 38 \\
(2 0 1) & & 1 & 26 & 1 & 33 & & 0\\
\hline \hline
\end{tabular}
\label{Table}
\end{table}

Table~\ref{Table} displays the calculated magnetic intensities for the reflections shown in Fig.~\ref{PeaksTemp} with the largest reflection in each column normalized to 100.  The reported intensities reflect the calculated magnetic cross section corrected for the resolution function of a triple-axis neutron spectrometer.  Calculated intensities are shown for Mn moments ordered in each of the four one-dimensional irreducible representations, as well as for Dy moments ordered in the $\Gamma_{2}$ representation.  Previous magnetization measurements on DYMO\cite{Harikrishnan2011} with $\vec{H}$~$||$~$\vec{c}$ revealed a low-temperature state with a spontaneous magnetization of $\approx$0.5~$\mu_{B}$ per formula unit (measured at 2~K).  Of the four irreducible representations present in hexagonal rare-earth manganites only the $\Gamma_{2}$ structure is consistent with either ferrimagnetism or ferromagnetism; undoped hexagonal DyMnO$_{3}$ is ferrimagnetic at low temperatures and is known to order in the $\Gamma_{2}$ representation.\cite{NandiDMO}  When ordering in the $\Gamma_{2}$ structure the magnetic cross sections of the ($H$~0~1) reflections depend only on the ordered moment of the Mn ions while those of the ($H$~0~0) reflections depend only on the ordered moment of the Dy ions.  The Mn ordered moment is therefore found to be almost temperature independent below 3.4~K, while the Dy ordered moment decreases continuously with increasing temperature between 2.1~K and 3.4~K.

\begin{figure}
\centering
\vspace{-5mm}
\includegraphics[width=9.3cm]{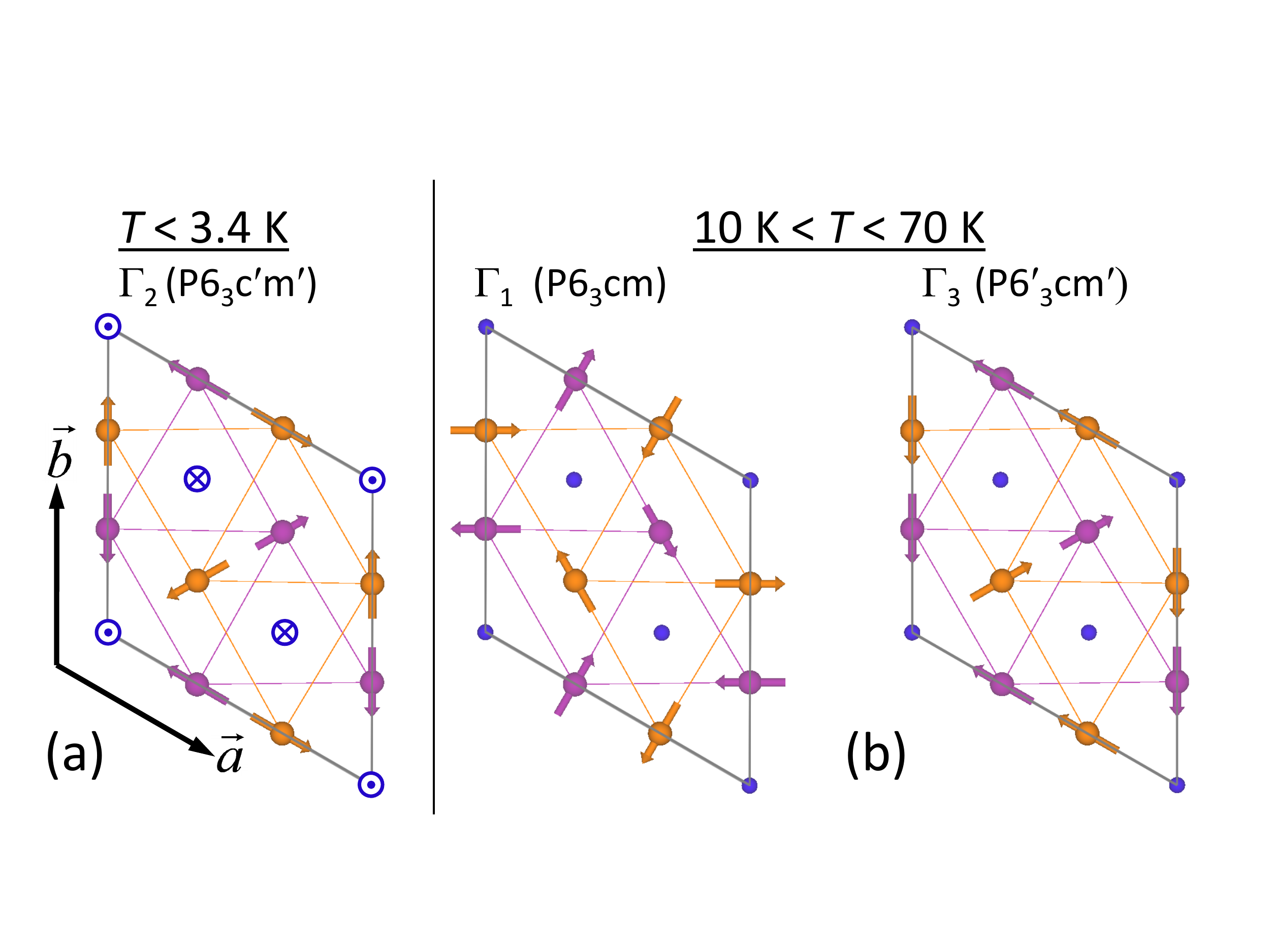} \vspace{-12mm}
\caption{(Color online) (a) Magnetic structure in the low-temperature ferrimagnetic phase, present below $T_{N}^{\textrm{Dy}}$~=~3.4~K.  The structure belongs to the $P6_{3}c^{\prime}m^{\prime}$ magnetic space group ($\Gamma_{2}$ irreducible representation).  Mn ions at $z$~=~0 are shown in purple, Mn ions at $z$~=~1/2 are shown in orange, and the rare-earth ions are shown in blue.  The refined ordered moments are 3.7~$\pm$~0.4~$\mu_{B}$ on the Mn ions and 3.1~$\pm$~0.3~$\mu_{B}$ on the Dy ions.  (b) The two possible magnetic structures for the antiferromagnetic phase: $P6_{3}cm$ magnetic space group ($\Gamma_{1}$ irreducible representation) on the left and $P6^{\prime}_{3}cm^{\prime}$ magnetic space group ($\Gamma_{3}$) on the right.  The Mn ordered moment is 3.5~$\pm$~0.4~$\mu_{B}$.  Any Dy ordered moment in this temperature range is too small to be measured by neutron diffraction.  The structural unit cell is outlined in gray} \vspace{-2mm}
\label{Order}
\end{figure}

Assuming a ferrimagnetic structure with ordered moments on the 2$a$ and 4$b$ sites that are antiparallel but equal in magnitude, a refinement of magnetic reflections measured at 1.6~K reveals ordered moments of 3.7~$\pm$~0.4~$\mu_{B}$ for the Mn ions and 3.1~$\pm$~0.3~$\mu_{B}$ for the Dy ions (1.6~$\pm$~0.2~$\mu_{B}$ for each rare-earth site).  This structure is displayed in Fig.~\ref{Order}(a).  (The crystal structures shown in Figure~\ref{Order} were produced using \emph{VESTA}.\cite{Momma})  When magnetic domains are fully aligned this ordered moment will lead to a net magnetization of 3.1~$\pm$~0.3~$\mu_{B}$ per unit cell (or 0.52~$\pm$~0.05~$\mu_{B}$ per formula unit) along the crystallographic $c$ axis, which is consistent with the reported bulk spontaneous magnetization.\cite{Harikrishnan2011}  Allowing for different ordered moments on the Dy $4b$ and $2a$ sites did not appreciably improve the fit of the 1.6~K data, nor could allowing for different moments on these sites produce a comparably good fit while yielding a net moment consistent with magnetization measurements.  A ferromagnetic ground state can likewise be excluded. A ferromagnetic state with equal ordered moments on the 2$a$ and 4$b$ sites would give no intensity for the ($H$~0~0) reflections; when ordering in the $\Gamma_{2}$ structure the intensities of these reflections depend on the difference in moment at the rare-earth sites: $I(\vec{Q}_{H00})~\propto~|\vec{M}_{4b}-\vec{M}_{2a}|^{2}$, where the ordered moments on the 2$a$ and 4$b$ sites are $\vec{M}_{2a}$ and $\vec{M}_{4b}$. A ferromagnetic state with different ordered moments on the 2$a$ and 4$b$ sites is consistent with the neutron diffraction data only while yielding a net moment inconsistent with magnetization measurements.  The Mn ordered moment at 1.6~K is close to the full moment value; however, the Dy ordered moment is considerably reduced from the full 10~$\mu_{B}$ value of the $^{6}H_{15/2}$ Dy$^{3+}$ ions.  The low temperature magnetization in undoped DyMnO$_{3}$ would suggest a similarly reduced ordered moment for the Dy ions in the ferrimagnetic phase.\cite{Ivanov2006}

The magnetic structure in the antiferromagnetic phase (10~K~$<$~$T$~$<$~70~K) is consistent with an ordered moment on only the Mn sublattice, with order according to either the $\Gamma_{1}$ or $\Gamma_{3}$ irreducible representation.  These magnetic structures are shown in Fig.~\ref{Order}(b).  It should be noted that neither of these possibilities are consistent with the structure of the Mn moments in (h)DyMnO$_{3}$, which order in the $\Gamma_{4}$ representation above the low-temperature ferrimagnetic phase.\cite{Wehrenfennig}  A refinement of magnetic reflections measured at 25~K gives an ordered moment of 3.5~$\pm$~0.4~$\mu_{B}$ for the Mn sublattice.  The ordered moment at 25~K is comparable in size to that determined at 1.6~K, suggesting a spin reorientation transition where Mn spins rotate between 3.4~K and 10~K with little change in magnitude.  Element specific x-ray resonant magnetic scattering measurements on hexagonal DyMnO$_{3}$\cite{NandiDMO} and HoMnO$_{3}$\cite{NandiHMO} have reported a weak rare-earth moment at comparable temperatures, induced by a splitting of the ground-state crystal field doublet.  While DYMO might similarly feature a weak Dy ordered moment in this temperature range, neutron scattering measurements will be sensitive to only the much larger ordered moment of the Mn sublattice and the refinement is not improved by allowing for an ordered moment on the Dy sublattice at 25~K.

The magnetic structure of YMnO$_{3}$ is known to be the $\Gamma_{3}$ representation.\cite{Fiebig2000}  If the antiferromagnetic phase of DYMO were likewise $\Gamma_{3}$, the spin reorientation transition into the low-temperature $\Gamma_{2}$ phase would differ from the spin reorientation transitions previously observed in hexagonal rare-earth manganites.  A magnetic structure in the $\Gamma_{1}$ irreducible representation can be transformed into the $\Gamma_{2}$ representation through a 90$^{\circ}$ counterclockwise rotation of all Mn moments.  However, a transformation of a structure in the $\Gamma_{3}$ representation as shown in Fig.~\ref{Order}(b) into the $\Gamma_{2}$ representation would require a 180$^{\circ}$ rotation of only the Mn moments in the $z$~=~$\frac{1}{2}$ plane.  In other hexagonal rare-earth manganites, it has been suggested\cite{Fiebig} that spin reorientation transitions between the four irreducible representations occur via in-phase or antiphase rotations where the moments in adjacent Mn layers rotate with an equal (in-phase) or opposite (antiphase) direction.  For example, the Mn moments in HoMnO$_{3}$ display an in-phase reorientation transition\cite{Vajk2005} ($\Gamma_{4}$ to $\Gamma_{3}$) at $T$~$\approx$~40~K and an antiphase transition ($\Gamma_{3}$ to $\Gamma_{1}$) at $T$~$\approx$~8~K, while the Mn moments in (h)DyMnO$_{3}$ display an antiphase rotation\cite{Wehrenfennig} ($\Gamma_{4}$ to $\Gamma_{2}$) at $T$~$\approx$~8~K.  Further, while materials such as ScMnO$_{3}$,\cite{Munoz2000} Er$_{x}$Y$_{1-x}$MnO$_{3}$,\cite{Sekhar} and YMnO$_{3}$ under pressure\cite{Kozlenko} have been reported to feature broad temperature ranges where the magnetic structure is not one of the four irreducible representations, the observed structures can still be described as an in-phase or antiphase rotation of all Mn moments away from one of the principal structures.

\subsection{Field dependence of the magnetic order}

\begin{figure}
\centering
\includegraphics[width=7.3cm]{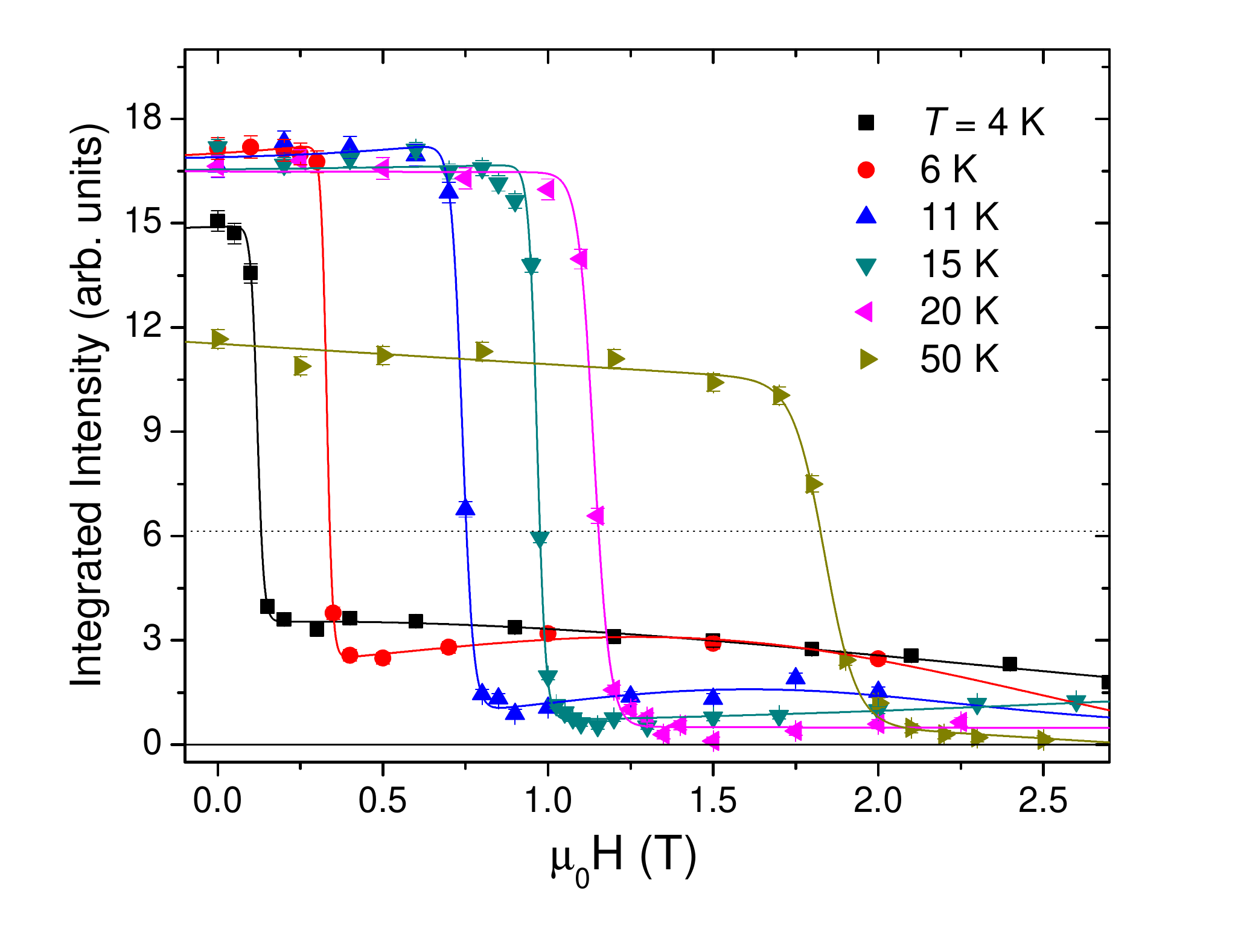} \vspace{-4mm}
\caption{(Color online) Integrated intensity of the (1~0~0) peak as a function of field.  The data were taken while increasing field.  Lines are a guide to the eye. The dotted horizonal line represents the expected intensity at 2~K in zero field.}\vspace{-2mm}
\label{Field1}
\end{figure}

Below $T_{N}^{\textrm{Dy}}$~=~3.4~K DYMO is ferrimagnetic with a spontaneous net magnetization.  Above $T_{N}^{\textrm{Dy}}$, $M(H)$ curves display symmetric magnetization steps at a critical field value that increases with temperature; the size of these steps decreases with increasing temperature, becoming immeasurably small around 40~K.\cite{Harikrishnan2011}  This behavior is remarkably similar to that previously reported in (h)DyMnO$_{3}$,\cite{Ivanov2006} but the moment increase associated with the magnetization steps in DYMO is about one half of the increase in (h)DyMnO$_{3}$.  Figure~\ref{Field1} displays the intensity of the (1~0~0) magnetic Bragg reflection as a function of field with $\vec{H}$~$||$~$\vec{c}$ at temperatures between 4~K and 50~K, measured while increasing the field after zero field cooling.  For each of these temperatures, we find a sudden drop in the (1~0~0) intensity at a critical field value that increases with temperature.  As was shown in Fig.~\ref{PeaksTemp}, the intensity of the (1~0~0) reflection will be much higher in the antiferromagnetic phase than in the low-temperature phase, such that this behavior is easily associated with a field-induced transition into the ferrimagnetic phase.  The data definitively show a spin reorientation of the Mn moments along with the field-induced ferrimagnetic ordering of the Dy moments, as the intensity of this reflection would not drop if the Mn moments remained ordered in the original structure.  The intensity of this reflection in the field-induced phase is weaker than in zero field at low temperature [the dotted horizontal line represents the expected intensity of the (1~0~0) reflection at 2~K and zero field].  This likely arises from a considerably smaller Dy ordered moment, consistent with the smaller magnitude of the magnetization steps at higher temperatures in the $M(H)$ curves.  Interestingly, neutron diffraction is sensitive enough that this transition is still clearly measurable in the 50~K data, while it could not be measured above 40~K in the magnetization data.\cite{Harikrishnan2011}

\begin{figure}
\centering
\vspace{-4mm}
\includegraphics[width=7.3cm]{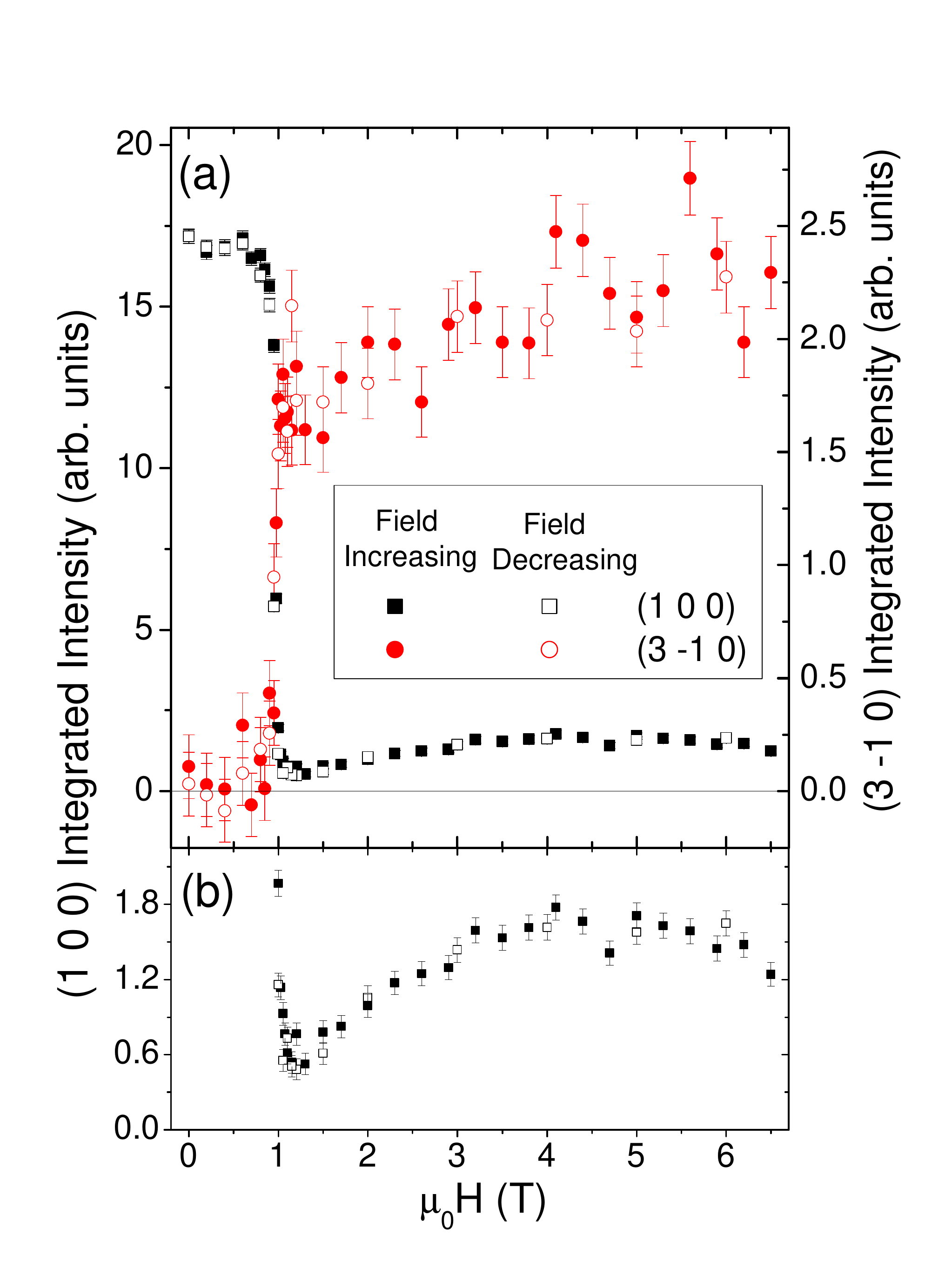} \vspace{-4mm}
\caption{(Color online) (a) Integrated intensities of the (1~0~0) (left scale) and (3~$\bar{1}$~0) (right scale) peaks measured at $T$~=~15~K.  Solid symbols represent measurements taken with increasing field, and open symbols represent those taken with a decreasing field. (b) Detail of the high field region for the (1~0~0) intensity.}\vspace{-2mm}
\label{Field2}
\end{figure}

The field dependence of the intensities at the (1~0~0) and (3~$\bar{1}$~0) reflections at $T$~=~15~K is displayed in Fig.~\ref{Field2}.  In Fig.~\ref{Field2}(a), both reflections show the expected transition at $\mu_{0}H_{c}$~$\approx$~1~T; data taken in rising and falling field show only a slight hysteresis, consistent with the small hysteresis of the magnetization steps in the bulk magnetization data.  In addition to DYMO, undoped (h)DyMnO$_{3}$ displays little hysteresis while magnetization steps in ErMnO$_{3}$\cite{Fiebig2001} display considerable hysteresis.  The (3~$\bar{1}$~0) reflection displays scattering in the low-temperature $\Gamma_{2}$ phase and very little scattering in the antiferromagnetic phase, such that this change in intensity is likewise consistent with a field-induced transition between the two.  A more detailed view of the (1~0~0) intensity in the high-field region is shown in Fig.~\ref{Field2}(b). Interestingly, the intensity of this peak rises slowly with increasing field up to about 4~T but then decreases with increasing field beyond 5~T.  This likely reflects the evolution of the system from ferrimagnetic to ferromagnetic as the field is increased, as the fully polarized state (ferromagnetic with an equal moment for the 4$a$ and 2$b$ sites) would display no scattering at (1~0~0).

\section{SUMMARY}

Below $T_{N}^{\textrm{Dy}}$~=~3.4~K Dy$_{0.5}$Y$_{0.5}$MnO$_{3}$ is magnetically ordered in the $\Gamma_{2}$ irreducible representation with antiferromagnetic Mn order and ferrimagnetic Dy order.  Between 3.4~K and 10~K we observe a spin reorientation transition of the Mn moments.  Antiferromagnetic order of the Mn sites persists up to $T_{N}^{\textrm{Mn}}$~=~70~K.  The magnetic structure of this phase is either $\Gamma_{1}$ or $\Gamma_{3}$, inconsistent with the $\Gamma_{4}$ order observed in undiluted (h)DyMnO$_{3}$.\cite{Wehrenfennig}  In the antiferromagnetic phase, a magnetic field applied parallel to the $c$ axis will drive a field-induced transition into the low-temperature ferrimagnetic phase with little hysteresis.  Further knowledge of the different magnetic structures and reorientation transitions displayed by the Mn moments in various (h)$R$MnO$_{3}$ materials should lead to improved understanding of the unusual 3$d$-4$f$ magnetic coupling present in these materials.\cite{Wehrenfennig}  While the magnetic structure of the antiferromagnetic phase cannot be definitively determined from this experiment, it is clear that Dy$_{0.5}$Y$_{0.5}$MnO$_{3}$ displays a zero-field spin reorientation transition not previously observed in any other hexagonal rare-earth manganite.  If the magnetic structure is $\Gamma_{3}$, then the $\Gamma_{3}$ to $\Gamma_{2}$ spin reorientation transition would involve only half of the Mn ions in a manner not observed in other (h)$R$MnO$_{3}$ materials.  The $H$-$T$ phase diagram of HoMnO$_{3}$ features a boundary between $\Gamma_{1}$ and $\Gamma_{2}$ phases\cite{Lorenz,Munawar} at $\mu_{0}H$~$\approx$~2~T, yet a zero field $\Gamma_{1}$ to $\Gamma_{2}$ spin reorientation transition has not been reported.  Further work will be necessary to distinguish between the $\Gamma_{1}$ and $\Gamma_{3}$ structures in the antiferromagnetic phase and to determine if the domain walls arising from this unique spin reorientation will display magnetoelectric coupling as is observed in compounds such as HoMnO$_{3}$\cite{Ueland,LottermoserPRB} or YMnO$_{3}$.\cite{Fiebig2002}

\section*{ACKNOWLEDGEMENTS}

We thank J.W. Lynn for guidance and helpful discussions.  J.S.H. acknowledges support from the NRC/NIST Postdoctoral Associateship
Program. This work was supported in part by the National Science Foundation under Agreement No. DMR-0944772.

* email: joel.helton@nist.gov
\bibliography{DYMO}
\end{document}